# Deep and diverse population synthesis for multi-person households using generative models with conditional inputs

**Hai Yang, Hongying Wu, Linfei Yuan, Xiyuan Ren, Joseph Y. J. Chow*, Jingqin Gao, Kaan Ozbay**

C2SMART Center

Department of Civil & Urban Engineering

New York University Tandon School of Engineering

* Corresponding author: joseph.chow@nyu.edu

## Abstract

Population synthesis is an increasingly important method used in numerous areas of transportation planning and analysis. Traditional methods such as iterative proportional fitting (IPF) produce stable and interpretable populations but cannot capture the interrelationships between household- and individual-level attributes. Recent deep learning methods offer this flexibility, yet can overfit high-dimensional attribute relationships without structural guidance and deviate from known structures. We develop a household level synthetic population generation framework that adapts the existing conditional input directed acyclic tabular generative adversarial network, or ciDATGAN, to multi person households. The framework combines household size specific data construction, directed acyclic graphs (DAG) informed dependency regularization, and conditional population inputs as deterministic anchoring to preserve intrahousehold associations. We apply the model to generate an open access synthetic population for New York State. The synthetic population includes nearly 20 million individuals and 7.5 million households in 2021. Validation against withheld benchmarks shows close agreement: joint distribution matching of Public Use Microdata Areas (PUMA), age, race, and disability against the full public use microdata sample (PUMS) yields an R-squared of 0.602 and a Jensen-Shannon distance of 0.280, pairwise Cramér's V differences between generated and benchmark records average below 0.007 at the person level, and classifier two-sample tests on complete household records yield AUC values of 0.527-0.549, close to chance level. The generated households reproduce cross member associations while increasing diversity by 10-17% over the PUMS sample and 13.2% over PopGen alone. A comparison with recent deep generative household synthesis studies indicates that these results are within the reported range while obtained at a larger scale and finer geographic resolution.

Associated data: https://zenodo.org/records/13732330

## 1. Introduction

Modern societies have become increasingly diverse and mobilized thanks to the rapid development of transportation innovations. Agent-based simulation models have become prevalent in evaluating policy and operation implementations with higher accuracies (Beckman et al., 1996; Arentze et al., 2007, Ye et al., 2009). The required activity inputs for these simulations are generated based on agents' demographic and socioeconomic attributes. For privacy protection, census bureaus release only small samples of detailed records together with aggregated marginals. For example, the public use microdata sample (PUMS) of the 2021 5-year American Community Survey (ACS) covers less than one million individuals in New York State (NYS), while the state population exceeds 20 million in 2021 (US Census, 2022). Therefore, proper population synthesis methods are required to upscale the limited sample to the full size.

Besides being used for agent-based models, a full-sized synthetic population dataset provides the opportunity to conduct social equity analysis more comprehensively. Equity metrics are mostly developed based on population segments (Martens et al., 2022). If just using marginal information, only a limited number of population segments can be defined for specific attributes and used to measure equity. A useful synthetic population should preserve the joint relationships between household attributes, person attributes, and the attributes of people who share a household. These relationships become difficult to estimate when many categorical variables are combined. The number of possible joint categories grows rapidly, and many combinations are sparse or absent in the survey sample. Independent person generation can match individual marginals but still create households with weak or unrealistic internal structure. Household synthesis therefore requires a model that can represent high order multivariate combinations without losing intrahousehold associations.

Synthetic population data are not new, and the most popular methodologies based on iterative proportional fitting (IPF) (e.g., Popgen (Krawczak et al., 2006)) have been around for many decades (Deming & Stephan, 1940). These methods are categorized as deterministic methods, and the synthetic population is constructed by replicating the input sample to match the marginal distributions (Farooq et al., 2013; Anderson et al., 2014; Casati et al., 2015). However, the multivariate distributions could become overly complex to derive when facing high-dimensional datasets. In addition, if certain existing population groups are not included in the sample, deterministic methods cannot recover the missing information, and probabilistic methods recover only limited diversity because the fitted distributions are tied to the observed sample.

Recently, machine learning (ML) techniques were introduced to address these scalability and accuracy issues, using neural networks to learn population distributions and attribute interdependencies, with generative adversarial networks (GAN) as the latest focus (Garrido et al., 2020; Badu-Marfo et al., 2022). Several have directly addressed joint household and person generation, showing a continuing tradeoff between generative flexibility, household consistency, user control, and agreement with geographic marginals (Aemmer and MacKenzie, 2022; Kukić et al., 2024; Sané et al., 2025; Qian et al., 2026). However, these fully learned generative models remain prone to overfitting in high-dimensional attribute spaces: as the joint categories of household and member attributes multiply while the sample stays fixed, the networks tend to memorize sparse observed combinations and amplify sample bias rather than learn the underlying dependency structure.

These strands of research reveal a persistent gap. Traditional procedures such as IPF and its extensions provide a dependable skeleton for the synthetic population because they reproduce known marginal totals, but they are not flexible enough to capture the interrelationships between household level and individual level attributes. Deep generative models possess the required flexibility, but when many categorical attributes interact they operate in a very large joint space, and without a reliable structural reference, or synthesis anchor, the learned relationships can drift away from known population structure. What is missing is a design that lets each family of methods do what it does best: the traditional method supplies the anchor, and the deep learning model supplies the flexibility.

In this study, we develop a household population synthesis framework around the existing conditional input directed acyclic tabular generative adversarial network (ciDATGAN) (Lederrey et al., 2022a). The contribution is the framework rather than a new GAN architecture. First, the framework converts household and person records into one household record with repeated member attribute blocks and trains a separate model for each household size. Second, it combines ciDATGAN with selected directed acyclic graphs (DAGs) and conditional population inputs. These components provide controls over conditional dependencies and geographic marginals. We apply the framework to New York State and release a synthetic population of nearly 20 million people and 7.5 million households as a public research resource. In essence, the framework couples a traditional deterministic synthesis as an anchor with a deep generative model, letting each compensate for the weakness of the other.

## 2. Literature review

Population synthesis is a technique widely used in fields such as transportation, urban planning, and epidemiology (Hafezi & Habibi, 2014; Salvini & Miller, 2005; Barrett et al., 2009; Auld & Mohammadian, 2010). The conventional approach employs IPF, a well-known statistical technique, along with probabilistic selection to generate synthetic populations. Sun et al. (2018) introduced a Bayesian network approach for population synthesis, showing improved results compared to traditional methods like IPF, with graphical representations for the structure. Saadi et al. (2016) developed a Hidden Markov Model (HMM)-based synthesis, demonstrating better performance than IPF with less data required; however, they did not match between households and individuals.

Deep generative models provide another way to estimate complex multivariate distributions. Borysov et al. (2019) introduced deep generative modeling based on Variational Autoencoders (VAE) to population synthesis, showing that VAEs handle high-dimensional data more effectively than Bayesian networks while maintaining the statistical properties of the original data. Comparing a Wasserstein Generative Adversarial Network (WGAN) and a VAE on a large-scale problem with over 60 variables and sampling zeros, Garrido et al. (2020) showed that WGAN outperforms VAE in generating valid synthetic attribute combinations while minimizing structural zeros. However, it produces less diverse populations.

These deep learning-based models have been shown to provide high-quality outputs compared to earlier deterministic and probabilistic methods. However, modelers have less control over the generation process (Lederrey et al., 2022b). The interdependencies among attributes are purely determined by the deep learning networks, which can lead to overfitting and be sensitive to sample bias. To overcome these limitations, Lederrey et al. (2022b) introduced the directed acyclic tabular GAN (DATGAN), which uses a data-specific DAG to regularize the generation process: the generator structure is restricted to follow the DAG, so only linked attributes are treated as correlated, limiting overfitting. Loss functions including standard GAN, Wasserstein GAN (WGAN), and WGAN with gradient penalty can be specified depending on the attribute types. Readers can refer to Lederrey et al. (2022b) for details. In multiple case studies, DATGAN outperformed other state-of-the-art deep learning models and proved more capable of producing low-probability categorical variables, which is important when the training sample is biased.

To further limit the potential influence of sample bias, Lederrey et al. (2022a) extended DATGAN into the conditional input DATGAN (ciDATGAN), in which the learned correlations are both restricted by the DAG and conditioned on selected attributes that are supplied externally during generation. If these conditional inputs are unbiased, the bias inherited from the training data can be significantly reduced. Figure 1 illustrates the training and generation processes of ciDATGAN. Readers can refer to Lederrey et al. (2022a, 2022b) for details on the model structure and case studies.

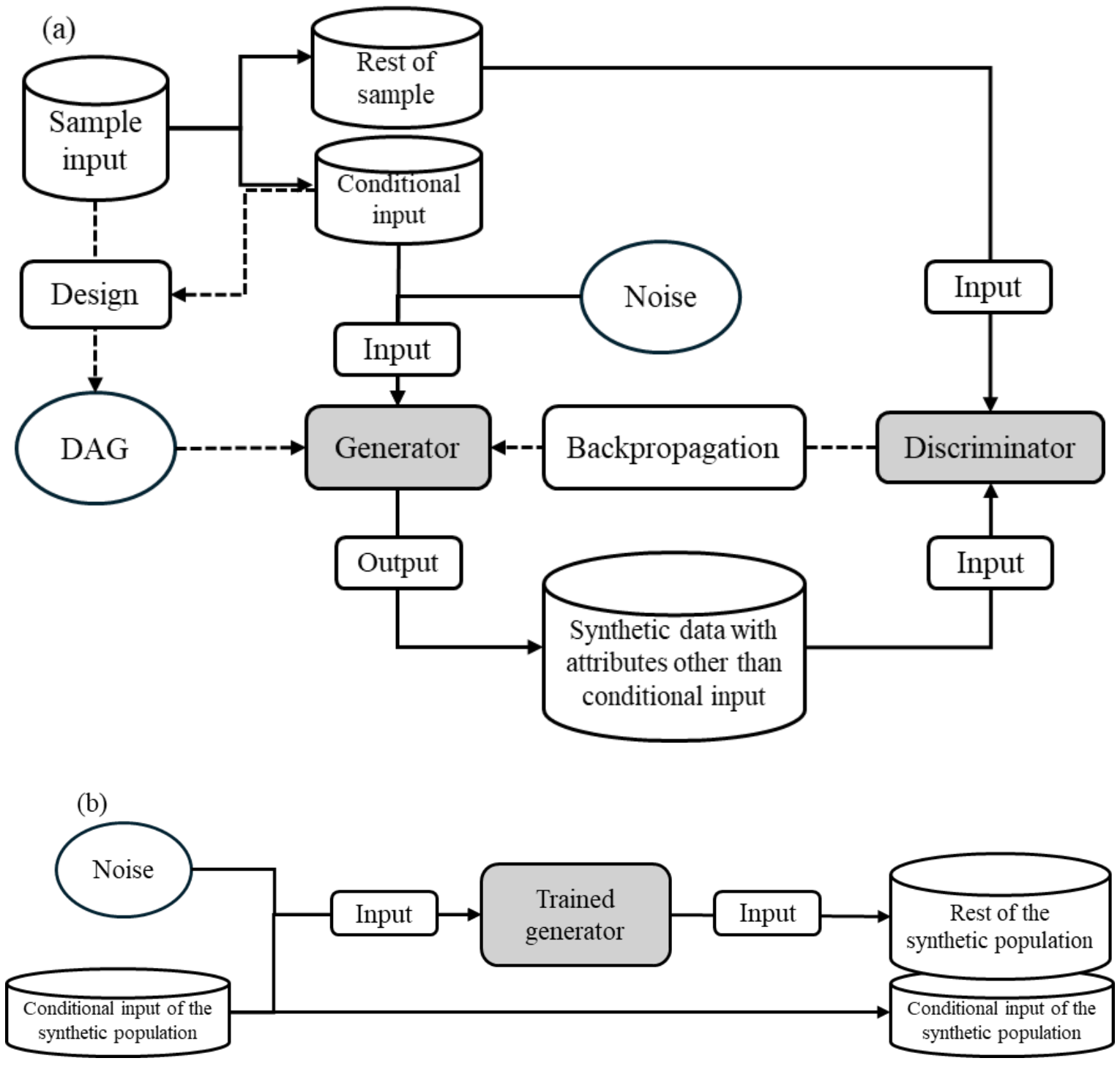

Figure 1. (a) Training process and (b) population generation process of ciDATGAN (adapted from Lederrey et al., 2022a).

Household synthesis adds a hierarchical structure to the population synthesis problem. Household attributes apply to all members, while each person also has individual attributes. Earlier multilevel approaches usually generate households first and then generate or assign individuals conditional on household characteristics (Anderson et al., 2014; Hu et al., 2018; Sun et al., 2018). This structure can preserve links between households and individuals, but cross member relationships may require additional matching or postprocessing. For example, Anderson et al. (2014) focus on the relationship between the household head and spouse.

Recent methods model household and person information more directly. Aemmer and MacKenzie (2022) use a household conditional variational autoencoder and an individual conditional variational autoencoder. Household outputs condition the generation of individuals, although the model does not directly learn all relationships among people in the same household. Kukić et al. (2024) propose one step Gibbs sampling for hierarchical records and show how consistency rules can prevent illogical households. Qian et al. (2026) place household and member attributes in a joint record and use transfer learning to match census tract marginals. Sané et al. (2025) also model household and person attributes jointly. Their MVAE Pop2 method trains separate variational autoencoders for different household sizes, while their alternative uses a single model with padded records.

Graph based synthetic data generation also appears outside household applications. Mattas et al. (2026) estimate a joint distribution for farm records with Bayesian network learning and combine it with nonparametric regression. The shared idea is that a graph can organize dependencies in a multivariate dataset. The present study uses the DAG differently. It regularizes the generator in ciDATGAN and is applied to records that contain a household and all of its included members. In our framework, the DAG does not independently define the complete synthetic distribution. It constrains the generator structure within ciDATGAN and is applied to records that contain household attributes together with indexed attributes for every included member. The main target is therefore the preservation of household to person and cross member associations.

Templ et al. (2017) and the simPop package represent a statistical approach to complex synthetic data. simPop provides methods including iterative proportional fitting, simulated annealing, logistic modeling, and data fusion. Our framework instead combines a conditional generative model with a DAG and an externally prepared conditional population. It also trains separate models by household size and evaluates intrahousehold member associations.

Among recent studies, the closest comparisons are the joint generative methods of Aemmer and MacKenzie (2022), Qian et al. (2026), and Sané et al. (2025), together with the household consistency approach of Kukić et al. (2024). The current framework adapts ciDATGAN through household size specific records, selected DAGs, and conditional population inputs. It then evaluates intrahousehold associations and applies the framework at the scale of New York State. The framework therefore combines ideas that have appeared separately in earlier studies. It uses a joint household record, as in recent deep generative household models. It separates models by household size, as in size specific VAE approaches. It also uses the DAG and conditional inputs of ciDATGAN to control dependencies and population totals.

The empirical contribution is to test this combination with household member associations and to generate a full state population for public use. Viewed against this literature, the distinguishing feature of the present framework is the explicit division of labor between a traditional deterministic synthesis and a deep generative model. Qian et al. (2026) decouple dependency learning from marginal alignment through transfer learning, and Kukić et al. (2024) obtain household consistency through the design of the conditional distributions. In our framework, an IPF anchored conditional population provides the structural skeleton, and the DAG regularized generator adds the flexibility needed to learn household to person and cross member relationships around that skeleton.

## 3. Methodology

We propose a population synthesis framework that involves both the deterministic model and ciDATGAN to generate households and corresponding personal synthetic populations (Figure 2). The framework include three main steps: (1) data preparation, (2) structural learning for DAG generation, and the (3) conditional input preparation for ciDATGAN.

The inputs are household records, person records linked to households, geographic controls, and selected conditional attributes. The output contains one record for each synthetic household and indexed attribute blocks for its members. The procedure has three stages. First, household and person data are converted into fixed length records for each household size. Second, candidate directed acyclic graphs are constructed and compared for each size group. Third, a conditional population is supplied to the trained ciDATGAN to generate the remaining attributes. These stages define the unit of generation, restrict the dependency structure, and align selected attributes with target population totals. The second and third stages embody the hybrid design of the framework: the third stage supplies a deterministic, IPF anchored skeleton, while the second stage equips the generative model with a structured yet flexible dependency space.

The initial household and person data may come directly from survey microdata or from a secondary synthesis at a finer geographic scale. The source does not change the three stages. In the New York application, the Public Use Microdata Sample (PUMS) provides the base records and PopGen supplies selected controls at the census tract level.

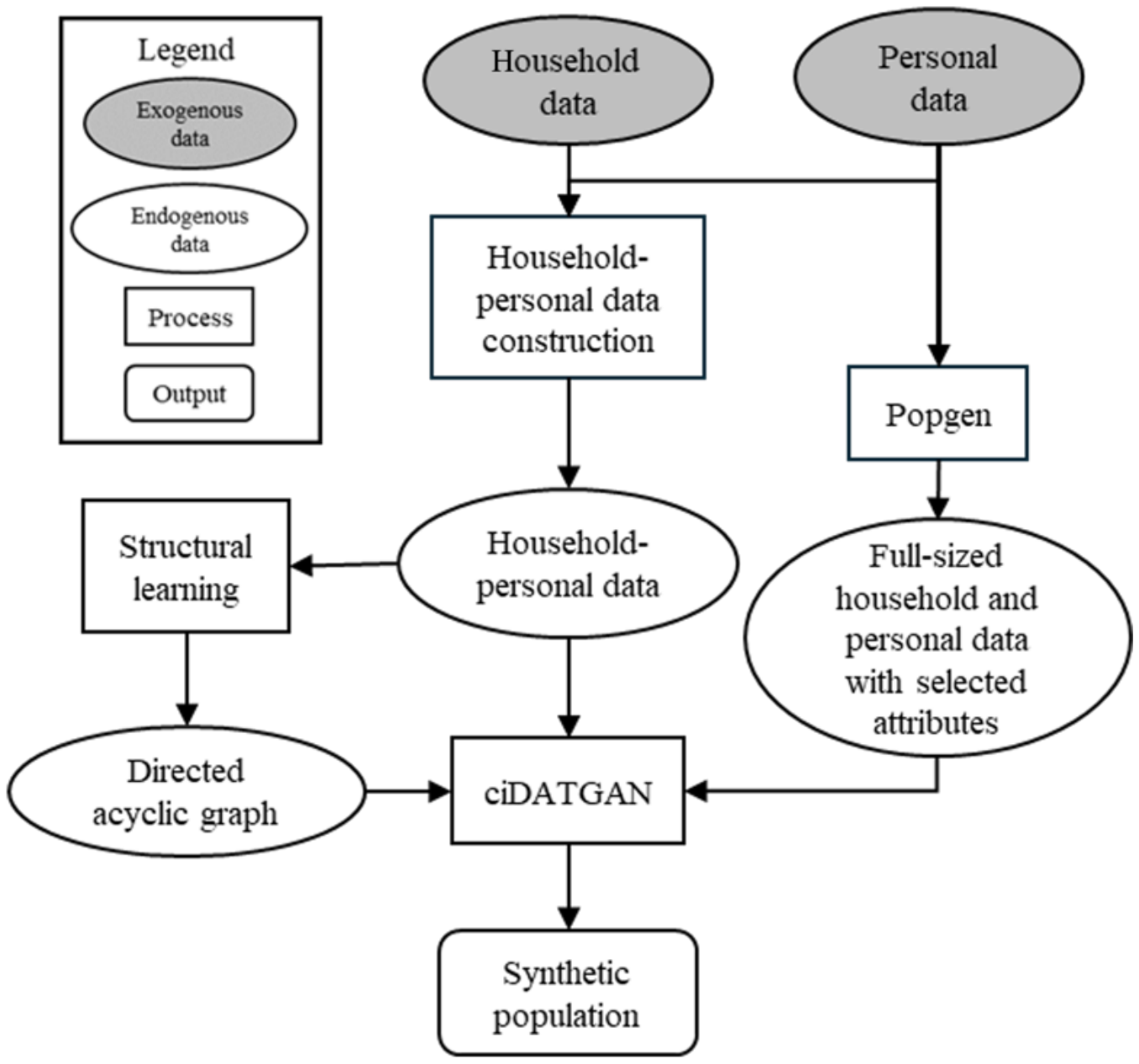

Figure 2. Flowchart of the population synthesis framework.

### 3.1. Household-personal data construction

Let $k$ denote household size. The household table contains one row for each household, and the person table contains one row for each member. The two tables are linked by household identifier. We first divide households into size groups. For a household of size k, the household attributes are followed by $k$ indexed person attribute blocks. Each row therefore represents one complete household. For example, the two age variables in a two-person household record are labeled AGEP_1 and AGEP_2. Figure 3 illustrates this construction.

One ciDATGAN model is trained for each household size. This is necessary because household size determines both the number of person blocks and the input dimension. A single record format cannot represent different household sizes with the same set of columns unless padding or another variable length design is introduced. We use household size rather than household type as the segmentation variable because size directly controls record dimension. Household type and relationship roles are not used to define the groups because the current data construction does not include a complete and consistent role classification. This choice supports direct estimation of cross member associations, but it does not impose explicit spouse or parent and child constraints. The use of household size as the segmentation variable follows precedents in the household synthesis literature, where Kukić et al. (2024) define a separate Gibbs sampler for each household size and Sané et al. (2025) train a separate variational autoencoder for each household size.

As household size increases, the number of attributes rises in proportion to the number of members, while the number of possible joint categories rises much faster. The framework therefore sets a maximum modeled household size $K$. Households with $k$ no greater than $K$ are generated with ciDATGAN. Larger households are completed with a deterministic procedure. $K$ is selected by considering population coverage, the number of records in each size group, record dimension, sparse combinations, and training time. In the New York application, $K$ equals five. This covers more than 95 percent of households and limits the largest model input to 48 attributes. The threshold is an application choice rather than a fixed property of ciDATGAN.

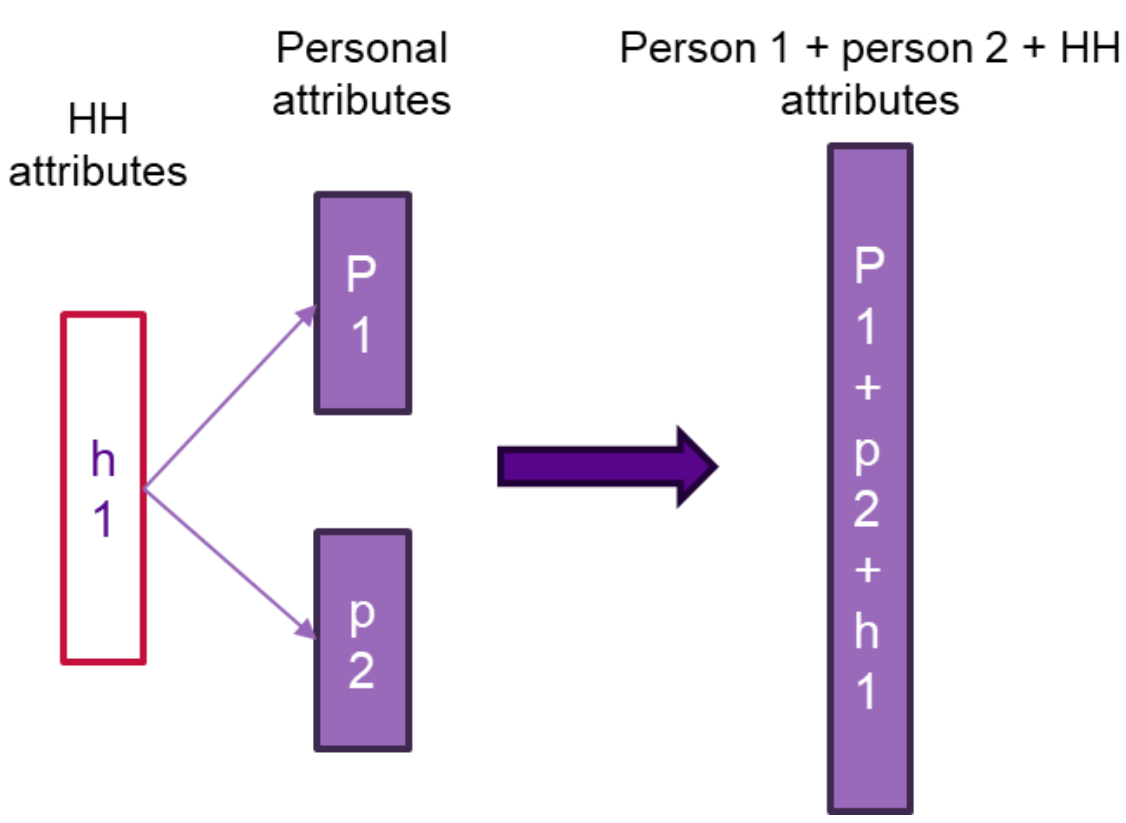

Figure 3. Integration of household and personal data for two-person household.

### 3.2. Structural learning for DAG generation

For each household size $k$, the DAG contains one node for every attribute in the household record. A directed edge identifies a permitted direct conditional dependency used by the generator. It does not establish a causal relationship. The graph regularizes ciDATGAN by limiting which variables can directly inform the generation of another variable (Lederrey et al., 2022b).

We consider six approaches to construct a candidate DAG for each household-size specific sample, and the best performing DAG is selected. Figure 4 illustrates the process. The Bayesian network, which is efficient in identifying dependency structures with prior knowledge (Heckerman, 2008), serves as the foundation for all methods. The DAG depicts conditional dependencies among variables (Ko et al., 2011), with edges indicating directed relationships and no cycles allowed (Pircalabelu et al., 2015). Bayesian networks are particularly suitable for large datasets with diverse data types (Yoo et al., 2014), and the "pgmpy" python package (Ankan & Panda, 2015) is chosen to facilitate data processing and model implementation for the Bayesian network.

The Hill Climbing algorithm is selected as the base for most methods, except for manual inputs in method 3, where edges are predefined. This algorithm explores the most promising paths for making changes to find optimal solutions (Selman & Gomes, 2006). The evaluation of these changes, which include the addition, removal, or reversal of edges, is based on a scoring system (Putri & Wijayanto, 2022). The process continues until no further improvement is possible. The Akaike Information Criterion (AIC) is used as the scoring method; it is preferred over the Bayesian Information Criterion (BIC) because it penalizes model complexity less (Vrieze, 2012). The Maximum Likelihood Estimator (MLE) is also involved to estimate model parameters. To further validate the model's robustness, cross-validation is used to assess model generalization and reduce overfitting. The mean and standard deviation of the AIC scores are calculated across five folds for each method.

The first approach, Focused Edges Based (FEB), constructs the DAG from a set of expert specified "focused edges" deemed essential. The second, Self Learning (SL), constructs the DAG entirely from self-learned edges optimized by Hill Climbing. The third, Human Adjusted Self Learning (HASL), manually adds the focused edges to the SL graph and removes repeated or cyclic edges. The fourth merges the FEB and SL graphs (FEB+SL). The fifth and sixth augment the focused edges with additional edges screened by machine learning, using Ordinary Least Squares in OLS Augmented Focused Edge (OLSAFE) and Random Forest in RF Augmented Focused Edge (RFAFE). Table 1 summarizes the six methods.

Table 1. Comparison of candidate DAG construction methods

| **Approach** | **Expert input** | **Added information** | **Role and limitation** |
| --- | --- | --- | --- |
| FEB | Required focused edges | None | Preserves selected dependencies. Results depend on the expert edge set. |
| SL | None | None | Provides a fully data driven reference. It can be sensitive to sparse samples and local optima. |
| HASL | Focused edges added after learning | Manual cycle resolution | Combines the learned graph with adjustment. Cycle resolution can affect reproducibility. |
| FEB + SL | Required focused edges | Union of the FEB and SL graphs | Uses both sources of structure. The union can produce a dense graph. |
| OLSAFE | Focused edges | OLS association screening | Adds linear screening. Numerical coding of categorical variables can affect coefficients. |
| RFAFE | Focused edges | Random Forest feature importance | Adds nonlinear screening. Results depend on the importance cutoff, and importance does not establish direction. |

The DAG with the best mean AIC score is then used as an input for ciDATGAN. Although other structure selection criteria could also be applied, such as BIC, held out likelihood, and Bayesian scoring methods, a comprehensive comparison of DAG selection criteria is beyond the scope of this study. The main objective is to develop and evaluate a household level population synthesis framework, rather than to determine which statistical criterion is universally most suitable for the DAG selection. AIC provides an established and consistent indicator for comparing candidate DAGs constructed from the same observations and variables, while considering both model fit and structural complexity. A systematic comparison of alternative DAG selection criteria and their effects on downstream synthesis performance is left for future research.

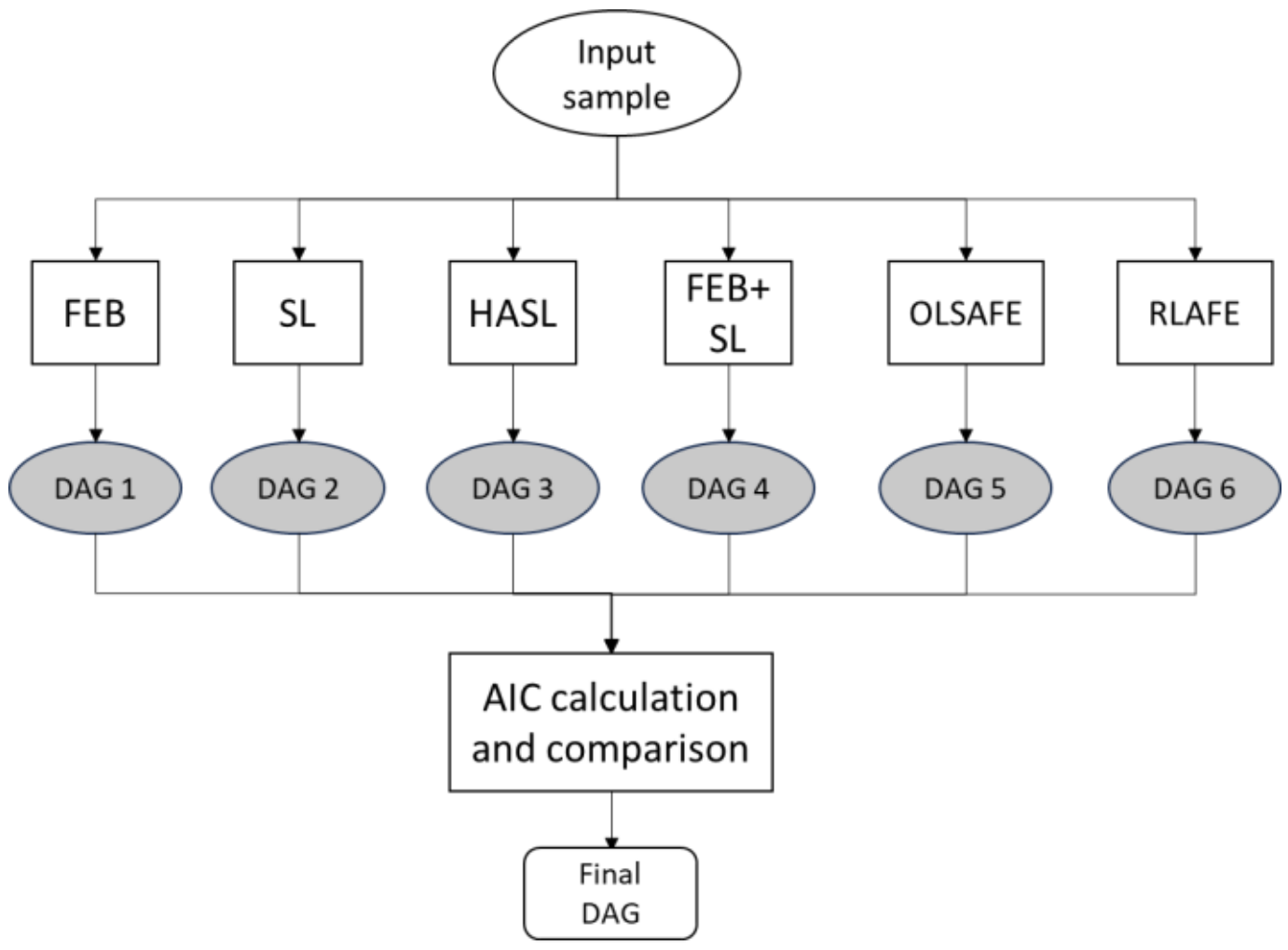

Figure 4. Flowchart of the structural learning for DAG generation.

### 3.3. Conditional input preparation for ciDATGAN

In addition to the trained ciDATGAN model and the learned DAG, population synthesis requires a separate population containing the conditional input attributes. These provided attributes act as roots, and the remaining attributes are generated around them using the interdependencies learned by ciDATGAN. The conditional input population must therefore have the same size as the final synthetic population, which calls for a separate, lower dimensional synthetic population.

Since the attributes serving as conditional input are relatively "fundamental" (e.g., age, gender), their misrepresentation in each municipal area of the survey sample is unlikely, and deterministic methods are suitable for synthesizing this lower dimensional full-sized population. We refer to the output as the conditional population, from which household size specific subsets are drawn for the final synthesis.

By combining the three procedures with ciDATGAN, we place substantially more control on the synthesis process than machine learning models alone: the adjusted inputs, DAGs, and conditional populations regularize associations at the individual, household to person, and member to member levels, while the generative core preserves the flexibility to synthesize new and more diverse samples than traditional methods. In other words, the deterministic conditional population serves as the skeleton of the synthetic population, anchoring the generation process to dependable marginals, while ciDATGAN provides the flexibility to learn the high order relationships that traditional methods cannot represent. This division of labor between the traditional component and the deep learning component is the central design principle of the proposed framework.

# 4. Application

## 4.1. Sample data

We apply the proposed framework to generate a full set of NYS synthetic population using the PUMS from 2021 5-year ACS (US census, 2024). The PUMS contains 956,365 individuals and 423,149 households from 2017 to 2021 after data cleaning. Though it covers the periods before and after COVID, the household level, person level, and cross member associations should remain consistent. To further reduce the bias, we prepare the conditional inputs as mentioned in Section 3.3 using the post-COVID information only.

The selected variables are listed in Table 2, with overly granular categories such as age and working industry (NAICS) aggregated. To capture spatial heterogeneity, we split PUMS into New York City (NYC) and non-NYC regions using the Public Use Microdata Areas (PUMAs).

Because NYC is the most densely populated and diverse region in the US, we assign the NYC PUMS from PUMA level to Census Tract (CT) level using Popgen and use the Popgen outputs as the input sample for the framework. Popgen has been widely used for agent-based simulation models (He et al., 2021) and reliably follows the input sample pattern. For the non-NYC population, however, Popgen fails to produce any output due to the excessive dimensionality, highlighting the scalability limits of traditional methods; we therefore reduce the number of attributes and categories (Table 2) to obtain CT-level information.

We then process the two population sets as described in Section 3.1, with a cut-off threshold of households with fewer than 6 members, forming 5 household-personal samples per region. Households with 6 or more people are replicated to complete the final synthetic population; they account for less than 5% of NYS households as in both regions, so the loss of information from replication is minor. Member attributes are re-labelled with member indices (e.g., “AGEP_1” and “AGEP_2” in a 2-person household). Figure 5 maps the distribution of households with 5 or fewer members.

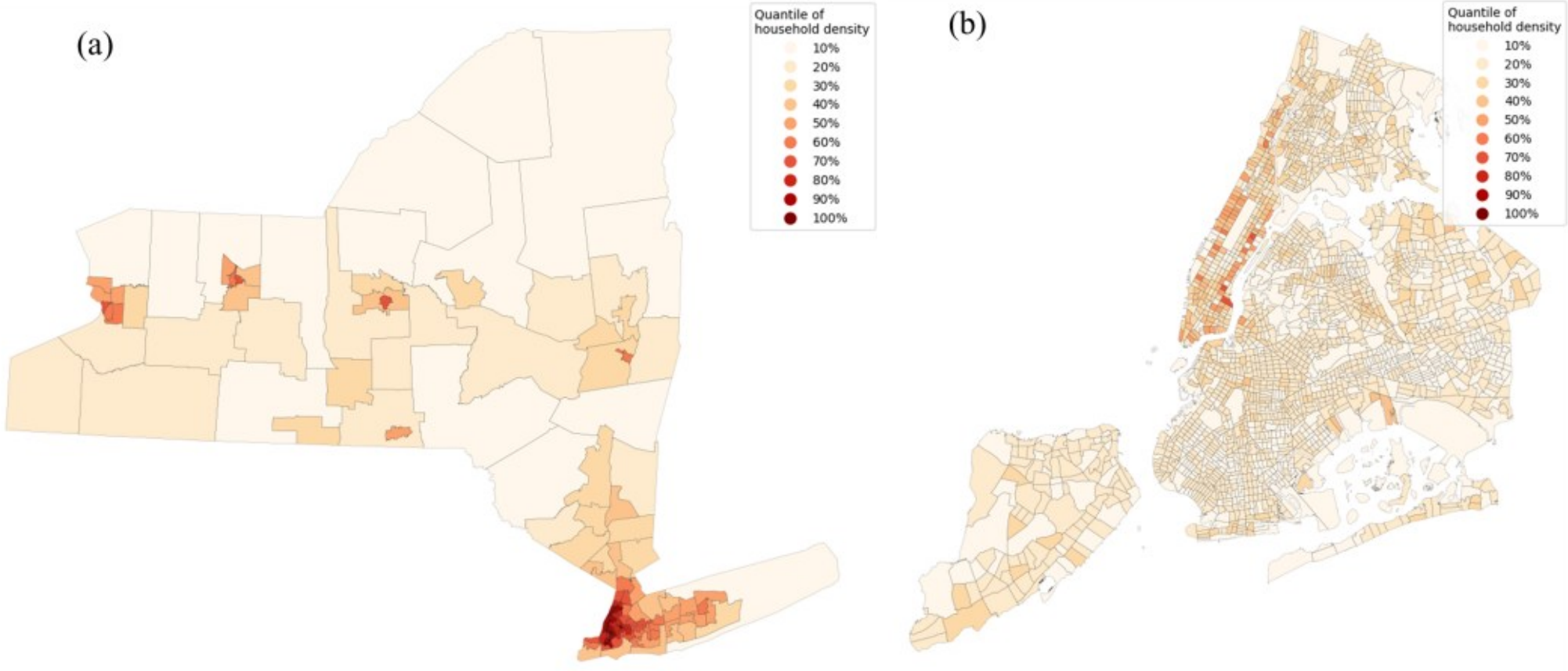

Figure 5. Density distribution of the households with less than or equal to 5 members in the input dataset in (a) NYS on PUMA level and (b) NYC on CT level

For non-NYC regions, the 1- to 5-person household samples contain 106,969, 87,253, 35,760, 28,921, and 12,320 observations with 12, 21, 30, 39, and 48 attributes, respectively. The corresponding NYC samples contain 56,184, 37,991, 20,797, 15,570, and 7,526 observations with 9, 15, 21, 27, and 33 attributes.

Table 2. Selected attributes of input samples

| | *Non-NYC region attribute (label name)* | *No. of values (range if continuous)* | *NYC region attribute (label name)* | *No. of values* |
|---|---|---|---|---|
| *Household attribute* | Residence area (PUMA) | 90 | Residence area (CT) | 2313 |
| | Income level (HINCP) | 9 | Income level (HINCP) | 9 |
| | Vehicle ownership (VEH) | 4 | Vehicle ownership (VEH) | 4 |
| *Personal attribute* | Age (AGEP) | 7 | Age (AGEP) | 7 |
| | English proficiency (ENG) | 5 | English proficiency (ENG) | 5 |
| | Commute trip length (JWMNP) | 0-140 min | Gender (SEX) | 2 |
| | Commute mode (JWTRNS) | 13 | Disability (DIS) | 2 |
| | School status (SCH) | 3 | Working industry (NAICSP) | 2 |
| | Gender (SEX) | 2 | Race white/non-white (RACWHT) | 2 |
| | Disability (DIS) | 2 | | |
| | Working industry (NAICSP) | 20 | | |
| | Race white/non-white (RACWHT) | 2 | | |

## 4.2. DAG results

Several edges deemed as essential are used as the base for DAG construction for the stated approaches. In this study, we fix the edges that connect residence areas with the ages of all household members, as well as the edges that connect residence areas with race. For example, for non-NYC 2-person households we fix the edges from "PUMA" to "AGEP_1", "AGEP_2", "RACWHT_1", and "RACWHT_2"; for NYC, "CT" substitutes for "PUMA". This ensures that age and racial distributions remain strongly associated with residential areas.

Table 3 summarizes the chosen best-performing DAG generation method among the six approaches for each household-personal dataset. Nearly all chosen methods are either HASL or FEB+SL, and their AIC scores are significantly higher than those of the other methods, highlighting the added value of human determined edges.

Table 3. Selected DAG generation approach

| | *Non-NYC population* | *NYC population* |
|---|---|---|
| *1-person household* | OLSAFE | FEB+SL |
| *2-person household* | FEB+SL | HASL |
| *3-person household* | FEB+SL | HASL |
| *4-person household* | FEB+SL | HASL |
| *5-person household* | FEB+SL | HASL |

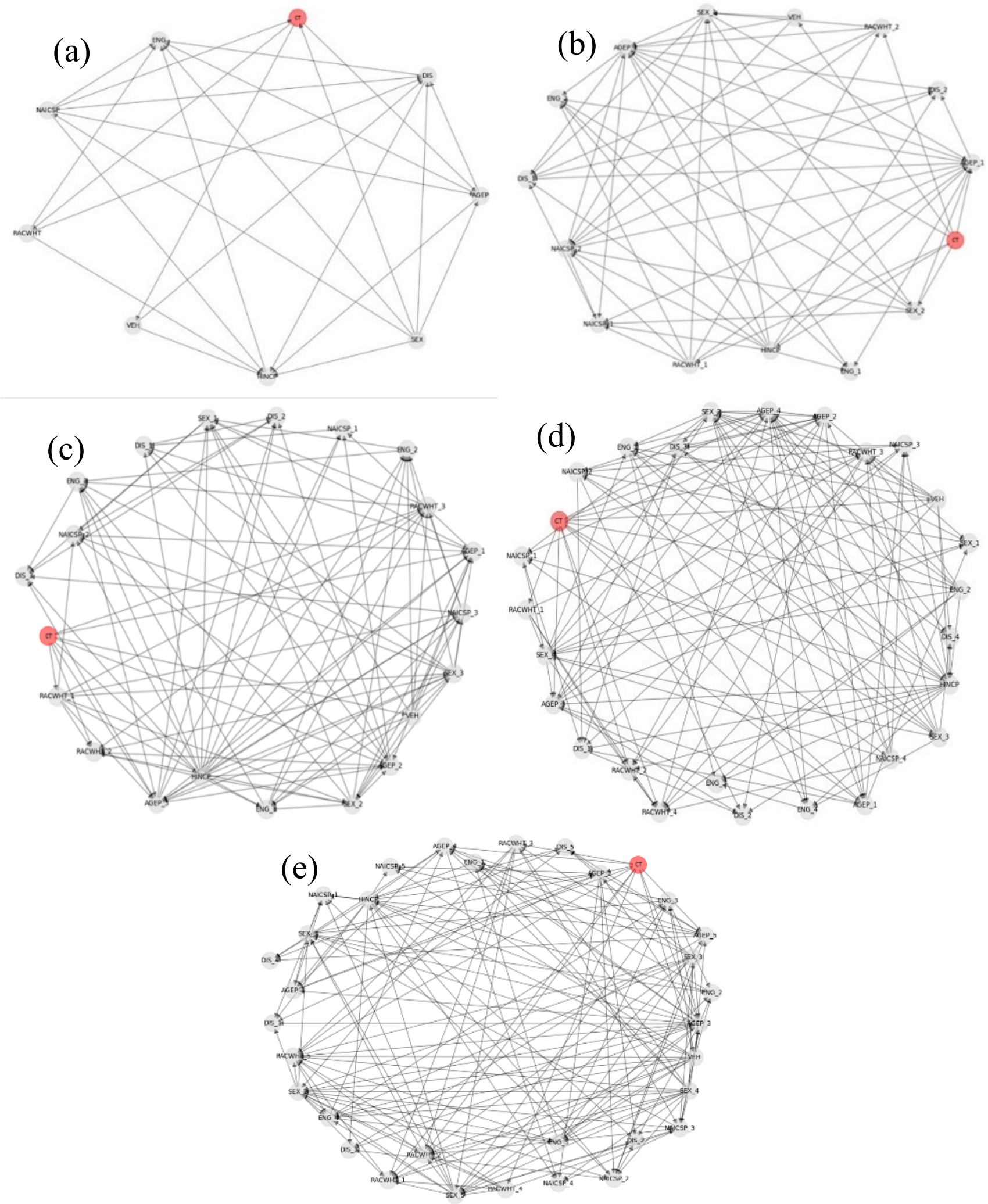

Figure 6. DAGs for (a) 1-person, (b) 2-person, (c) 3-person, (d) 4-person, and (e) 5-person households in NYC.

Figure 6 and Figure 7 illustrate the DAGs constructed by the selected methods for NYC and non-NYC populations. The DAG complexity grows roughly linearly with the number of attributes, from 25 links for the non-NYC 2-person households to 101 links for the 5-person households; this restricted growth reduces the computational burden of ciDATGAN training while maintaining the core associations.

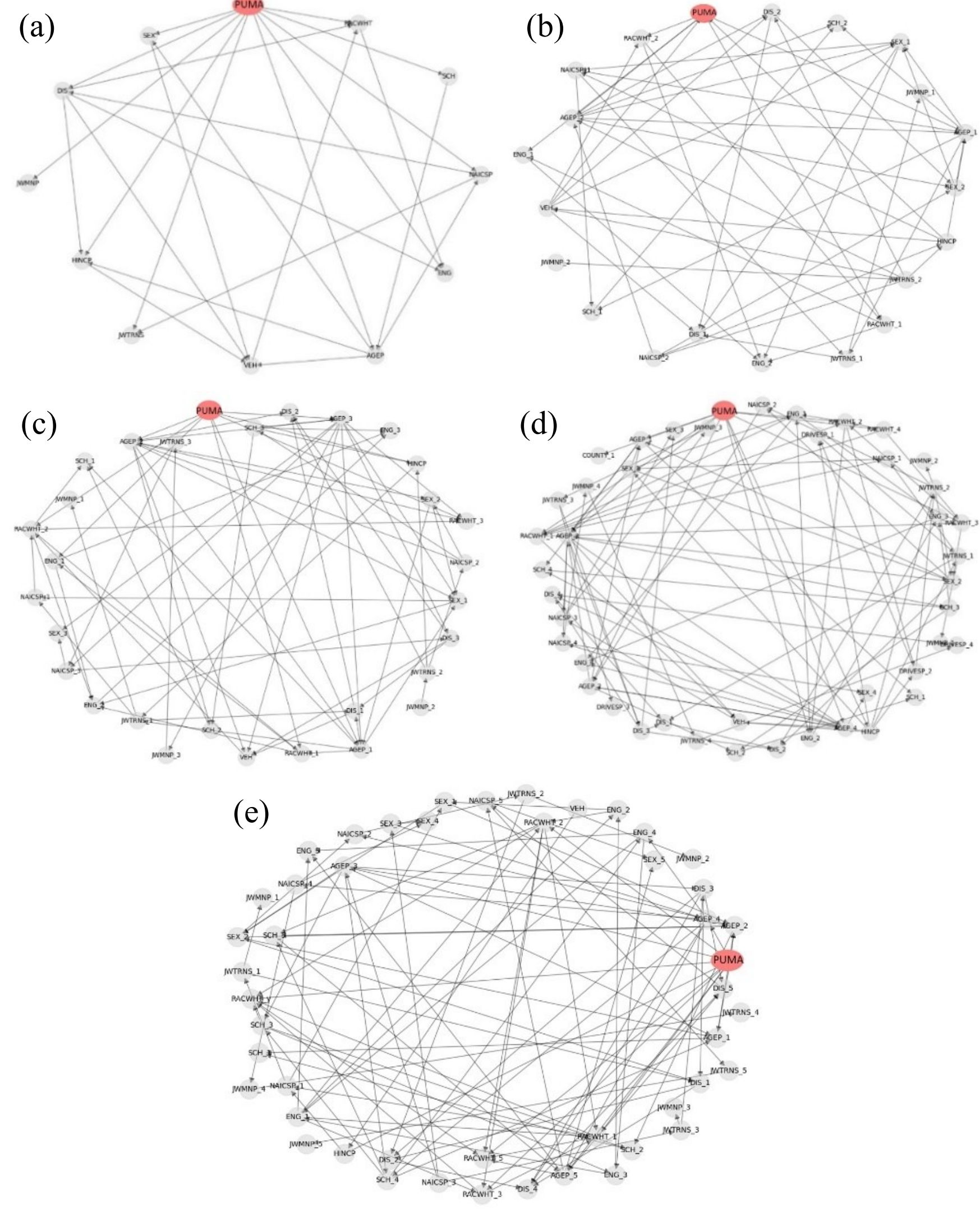

Figure 7. DAGs for (a) 1-person, (b) 2-person, (c) 3-person, (d) 4-person, and (e) 5-person households in the non-NYC region.

### 4.3. ciDATGAN parameters and conditional input

For each household size specific population, we train the ciDATGAN using the processed sample and corresponding DAG. As listed in Table 2, the majority of the attributes in both non-NYC and NYC populations are categorical. As suggested in Garrido et al. (2020) and Lederrey et al. (2022b), the Wasserstein loss function is particularly suitable for handling categorical information. Therefore, we specify the WGAN as the loss function. Readers can refer to Lederrey et al. (2022b) for the details of the embedded loss functions in ciDATGAN. We apply the same parameters recommended in Lederrey et al. (2022a) for the ciDATGAN training.

The conditional input attributes are residence area, age, and race, aligning with the "fixed edges" defined in Section 4.2 and bounding the synthetic population by these foundational attributes.

Following Section 3.3, we use Popgen to generate the conditional populations for non-NYC and NYC residents (Fig. 2), each containing the three attributes and matching the target population size from the 2021 ACS (US census, 2022).

Training and generation are run under Windows 11 with an i9-12900h CPU and a mobile RTX 3070Ti GPU. Ten sessions are conducted, each using 40% of the corresponding dataset for training, with a total run time of 14.25 hours; the NYC 5-person model takes the longest at 7.5 hours, and training occupies over 95% of each session. The trained models can be reused with updated conditional populations, and generation takes less than 10 minutes per population segment, making the framework more efficient than traditional methods for large-scale multivariate data.

## 4.4. Result validation

To validate the method, we evaluate the 100% population data based on the outcomes of the training using only the 40% data. The generated synthetic population contains 10,775,790 individuals and 4,267,578 households in non-NYC regions, and it includes 8,347,978 individuals and 3,283,033 households in NYC. The validation is organized in five layers: Section 4.4.1 compares selected marginals with census benchmarks; Section 4.4.2 evaluates the fitness and closeness of multivariate joint distributions; Section 4.4.3 validates person level associations; Section 4.4.4 validates household member associations; and Section 4.4.5 tests the temporal transferability of the trained models. Five complementary metric families are used across these layers: the coefficient of determination (R-squared) for the overall fitness of joint distribution comparisons, SRMSE and JSD for distributional closeness, Cramér's V for pairwise association strength, classifier two-sample test statistics for the separability of complete records, and plausibility checks for structural validity.

### *4.4.1. Comparison with census marginals*

We select four PUMAs, representing Albany, Rochester, the Upper East Side in Manhattan, and Buffalo, to showcase marginals in areas with distinct socio-demographic structures, aggregating NYC census tract information to PUMA level for comparability. Figure 8 illustrates marginal distributions for race, disability, disability and gender, race and disability, and age and disability; Figure 9 shows household vehicle ownership and household size. The number in each label indicates the category. Overall, individual attribute marginals closely match the census, while household marginals match less closely but preserve the general patterns. For joint labels, the notation lists the attribute codes with their category values. For example, the label "VEH 0 NP 1" denotes households owning no vehicle and containing one member.

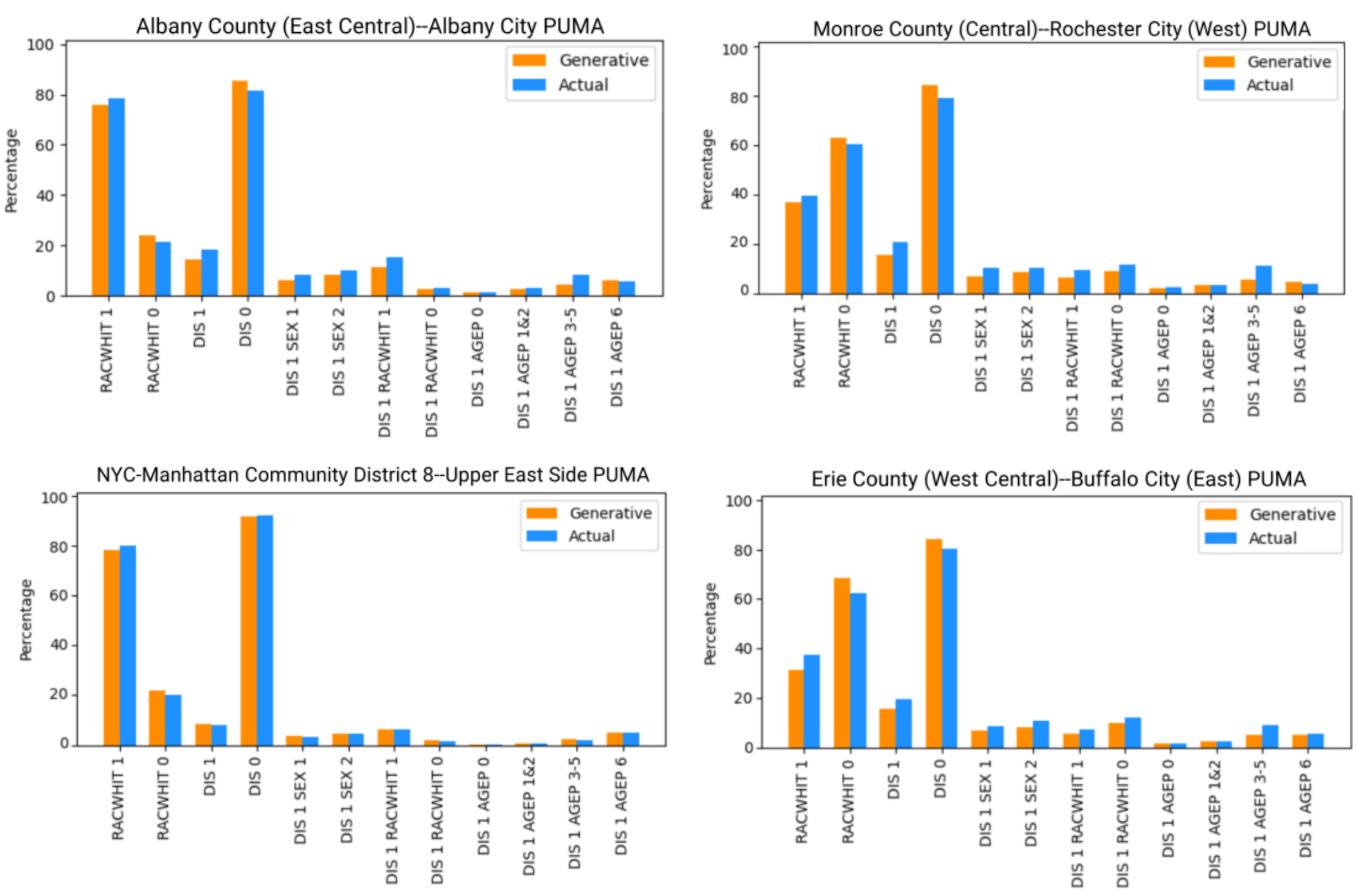

Figure 8. Marginal distributions for attributes of age, race, and disability in selected PUMAs.

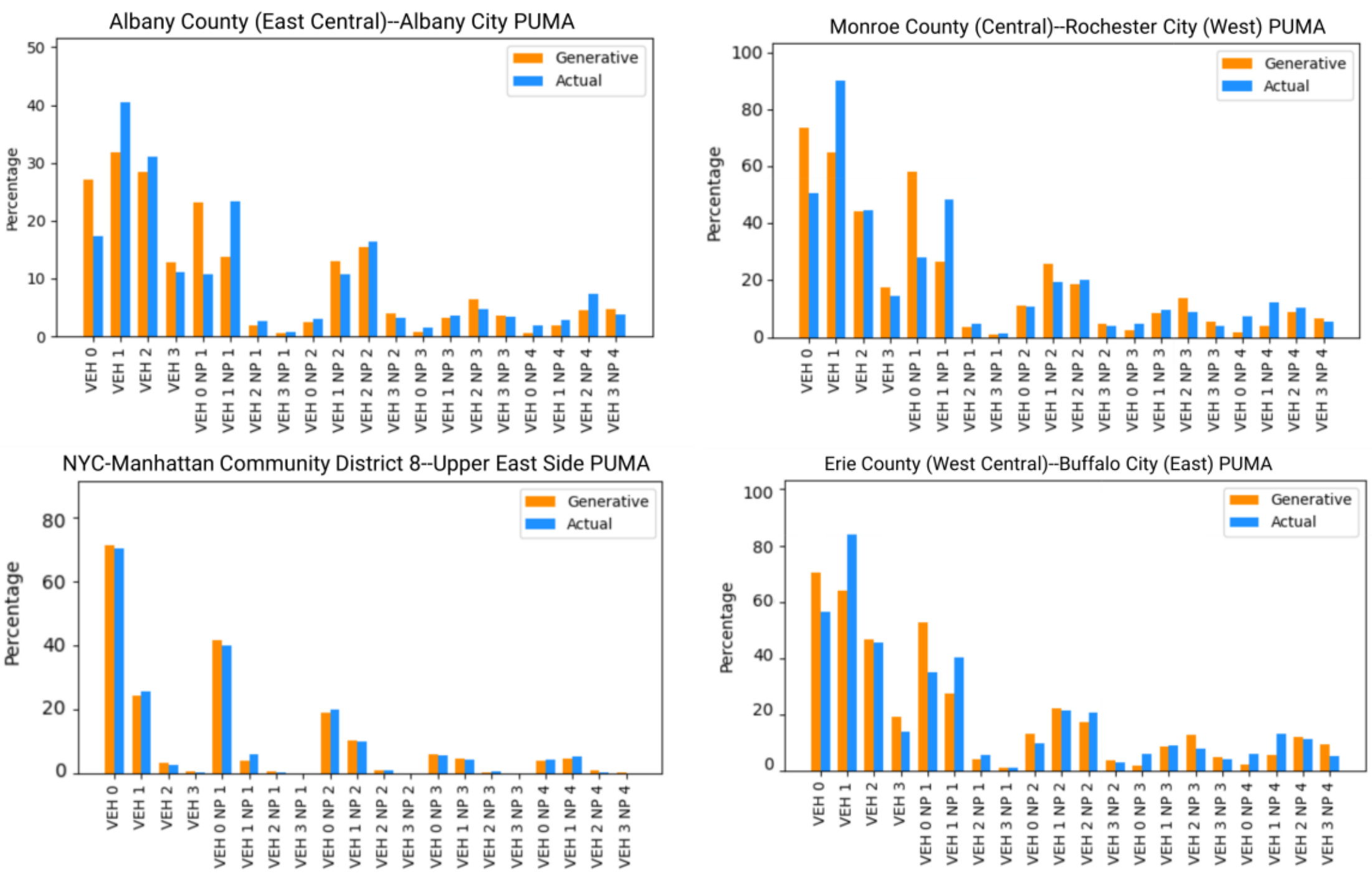

Figure 9. Marginal distribution for household attributes of vehicle ownership and household size in selected PUMAs.

We further compare marginals involving household income and working status, assigning all members of a household the same income level and treating the PUMS marginal as the census truth. Figure 10 shows the NYC-level marginals for income and employment status combined, with close resemblance across all groups. Figure 11 further breaks down income and working industry outside NYC; the synthetic pattern closely follows PUMS across all income-industry groups. Because the non-NYC income and industry cross tabulation contains many sparse categories, Figure 11 is best read as a coarse pattern comparison, and the quantitative assessment of such high order combinations is provided by the metrics reported in the remainder of this section.

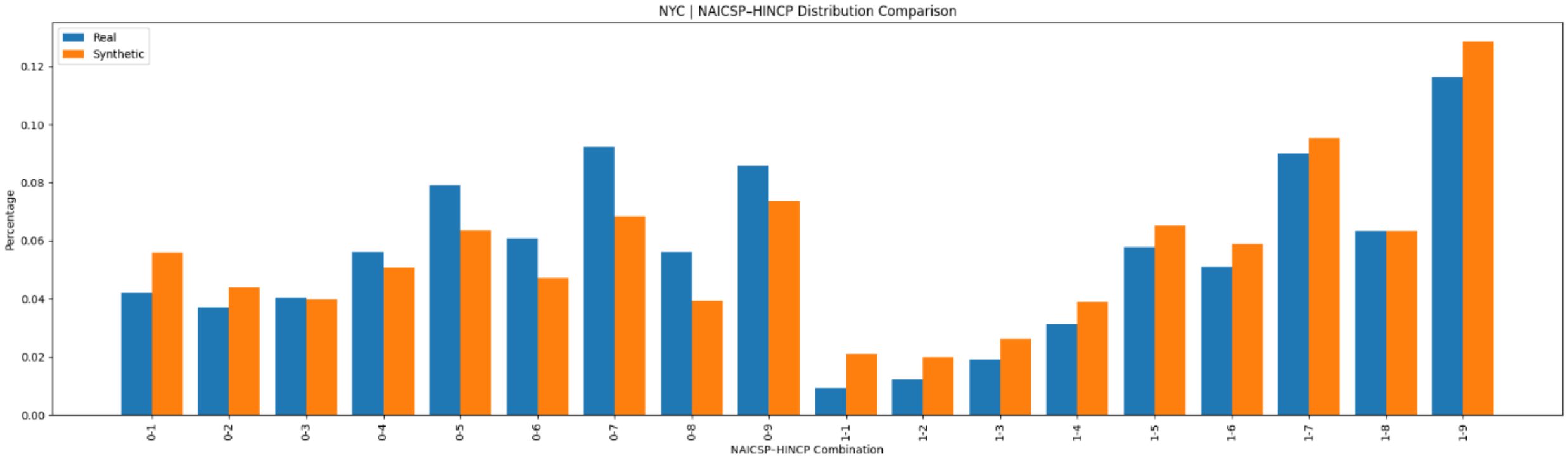

Figure 10. Marginal distribution for employment status and income-level in NYC.

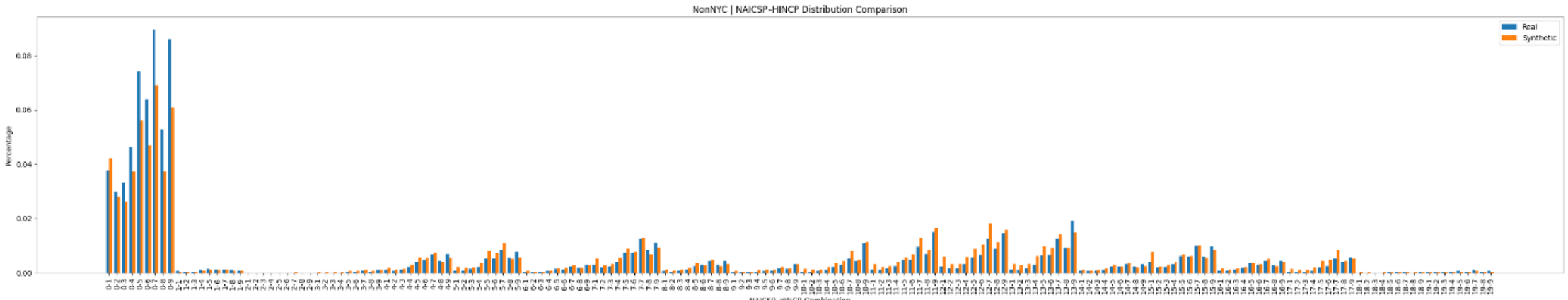

Figure 11. Marginal distribution for employment status and income-level in non-NYC area.

#### *4.4.2. Marginal fitness and closeness*

Because higher dimensional joint marginals are not available from the census, we compare multivariate joint distributions between the synthetic population and the 100% PUMS, beginning with residential area, age, race, and disability in Figure 12. Each point represents a unique joint category combination. The horizontal axis represents the proportion of that category combination in the benchmark data, while the vertical axis represents its proportion in the synthetic data. A point on the diagonal reference line indicates exact agreement. The distance from the diagonal indicates the size of the discrepancy. The greater dispersion near the origin arises because many high order combinations occur very infrequently in the benchmark data. We use R-squared to measure the overall fitness. We also use the Standardized Root Mean Squared Error (SRMSE) (Knudsen & Fotheringham, 1986) metric and the Jenson-Shannon distance (JSD) (Kim et al., 2022) to quantitively measure the closeness between the synthetic population and PUMS. The SRMSE is calculated using Eq. (1).

$$SRMSE = \frac{\sqrt{\frac{1}{N}\sum(\hat{x}_i - x_i)^2}}{\frac{1}{N}\sum x_i} \tag{1}$$

where $\hat{x}_i$ and $x_i$ represent the percentage of observations of population group $i$ in the synthetic population and PUMS. The population group $i$ is defined by the number of included attributes. Instead of calculating the absolute difference between each population group, the JSD uses entropy to measure the divergence between the two datasets, which is obtained using Eqs. (2) and (3).

$$JSD(\hat{X}|X) = \sqrt{0.5\left(KL\left(\hat{X}\middle|0.5(\hat{X}+X)\right) + KL\left(X\middle|0.5(\hat{X}+X)\right)\right)} \tag{2}$$

$$KL(\hat{X}|X) = \sum\hat{x}_i \log\left(\frac{\hat{x}_i}{x_i}\right) \tag{3}$$

The $KL$ in Eq. (3) stands for the Kullback-Leibler divergence. $\hat{X}$ and $X$ represent the full set of $\hat{x}_i$ and $x_i$. The SRMSE and JSD results are also presented.

The overall fitness visible in the joint distribution scatter plots is quantified with the coefficient of determination, computed over the proportions of all joint categories as Eq. (4).

$$R^2 = 1 - \frac{\sum(\hat{x}_i - x_i)^2}{\sum(x_i - \overline{x})^2} \tag{4}$$

where the summations run over all joint categories $i$ and the bar denotes the mean of the benchmark proportions. An R-squared value of one indicates that every joint category is reproduced exactly, while lower values indicate growing dispersion around the diagonal reference line.

Figure 13 expands the analysis to wider individual level combinations and to household level attributes. The fit for the personal attributes drops noticeably, pointing to the difficulty of such high order cross-tabulations, yet the R-squared level remains promising for a dataset of this size compared with previous studies.

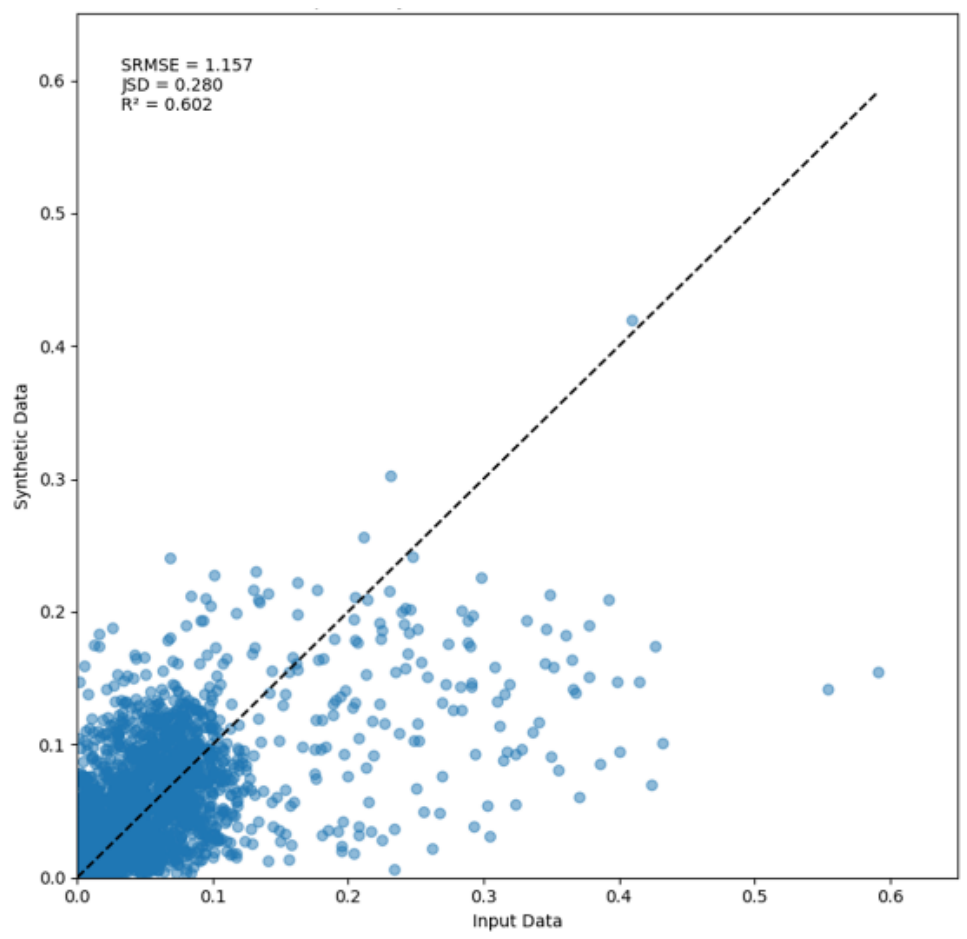

Figure 12. Joint distribution matching for combination of PUMA, age, race, and disability.

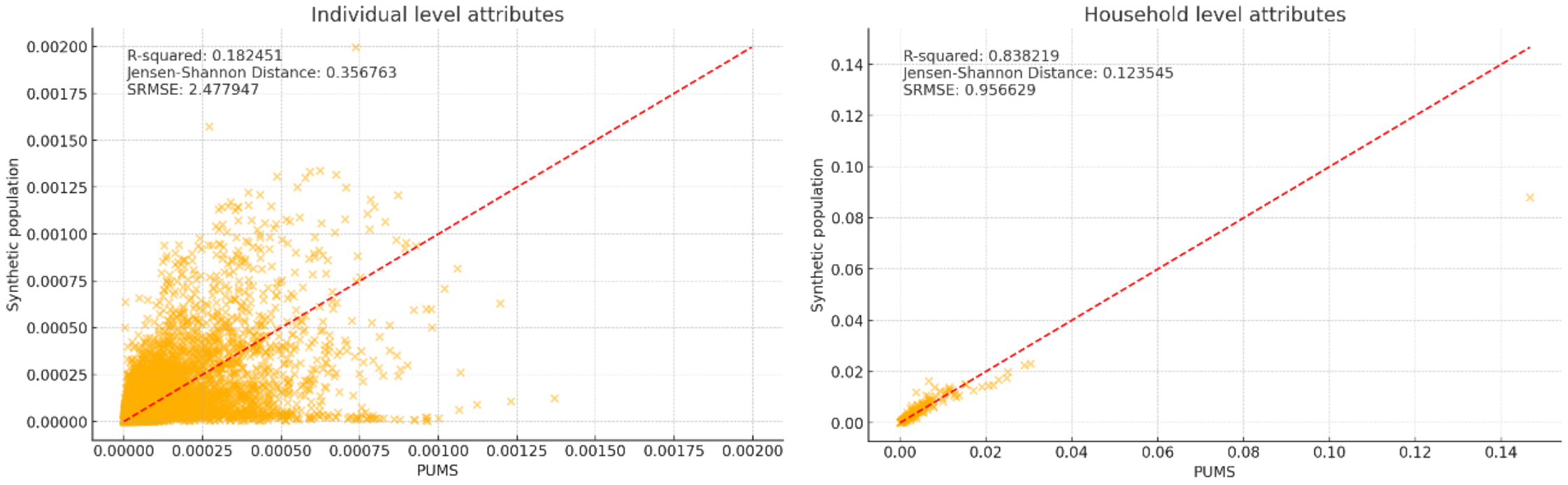

Figure 13. Joint distribution matching for combination of PUMA, age, race, gender, English proficiency, and disability (left); and combination of PUMA, vehicle ownership, number of household members, and household income level (right).

### *4.4.3. Personal level validation*

To extend the validation, we compare the generated NYC person records with a deterministic 20% held out PopGen sample. Pairwise dependencies are evaluated by calculating Cramér's V (Cramér, 1999) for all 15 pairs of common attributes. As shown in Table 4, the mean absolute difference between the generated and benchmark values is 0.0067, with a maximum of 0.0121. The strongest benchmark associations are age with working industry, age with disability, and disability with working industry. These patterns are closely reproduced in the generated data. The largest deviations occur for sex with disability, disability with race, and English proficiency with sex, but each absolute difference remains close to 0.01.

Cramér's V measures the strength of association between two categorical attributes. For a pair of attributes with r and c categories, $V$ is shown in Eq. (5).

$$V = \sqrt{\frac{\chi^2}{N \cdot min(r-1,\ c-1)}} \tag{5}$$

where $\chi^2$ is the chi-square statistic of the r × c contingency table between the two attributes and N is the number of records. $V$ ranges from zero under independence to one under perfect association. Dependency preservation is evaluated by comparing the $V$ value of each attribute pair in the generated data with the corresponding value in the benchmark data.

Table 4. Pairwise Cramér's V comparison for common personal attributes

| *Variable pair* | *Generated V* | *Benchmark V* | *Absolute difference* |
|---|---|---|---|
| *SEX and DIS* | 0.0244 | 0.0123 | 0.0121 |
| *DIS and RACWHT* | 0.0166 | 0.0281 | 0.0115 |
| *ENG and SEX* | 0.0215 | 0.0317 | 0.0101 |
| *ENG and NAICSP* | 0.0351 | 0.0448 | 0.0097 |
| *AGEP and DIS* | 0.3037 | 0.2942 | 0.0095 |
| *ENG and DIS* | 0.0947 | 0.0854 | 0.0093 |
| *AGEP and SEX* | 0.0629 | 0.0538 | 0.0091 |
| *AGEP and ENG* | 0.1186 | 0.1267 | 0.0080 |
| *DIS and NAICSP* | 0.1842 | 0.1915 | 0.0073 |
| *ENG and RACWHT* | 0.0203 | 0.0247 | 0.0044 |
| *AGEP and NAICSP* | 0.6397 | 0.6438 | 0.0041 |
| *SEX and NAICSP* | 0.0439 | 0.0419 | 0.0020 |
| *NAICSP and RACWHT* | 0.0221 | 0.0209 | 0.0012 |
| *SEX and RACWHT* | 0.0008 | 0.0019 | 0.0011 |
| *AGEP and RACWHT* | 0.0313 | 0.0309 | 0.0003 |

Following the classifier two sample test logic (Lopez-Paz & Oquab, 2016), we also evaluate whether the full joint pattern of the six personal attributes can distinguish generated records from benchmark records. We use a smoothed empirical likelihood ratio over the 448 observed joint patterns. The training sample contains 5,841,619 generated records and 1,168,403 benchmark records, while the test sample contains 2,504,794 and 501,736 records, respectively. As shown in Table 5, the area under the receiver operating characteristic curve is 0.5273, balanced accuracy is 0.52, and the Brier score is 0.25. These values are only slightly above chance, indicating limited but nonzero separability between the two datasets.

In the classifier two-sample test, a probabilistic classifier is trained to distinguish generated records from benchmark records and is evaluated on held out records that were not used for its training. The area under the receiver operating characteristic curve (AUC) equals the probability that a randomly selected generated record receives a higher predicted score than a randomly selected benchmark record; a value of 0.5 corresponds to chance level and indicates that the two datasets are indistinguishable to the classifier. Balanced accuracy ($BA$) averages the true positive rate and the true negative rate, as shown in Eq. (6).

$$BA = \frac{1}{2}\left(\frac{TP}{TP + FN} + \frac{TN}{TN + FP}\right) \tag{6}$$

where TP, TN, FP, and FN denote true positives, true negatives, false positives, and false negatives at the 0.5 decision threshold. The Brier score measures the calibration of the predicted probabilities, as defined in Eq. (7).

$$BS = \frac{1}{N_c}\sum(p_i - y_i)^2 \tag{7}$$

where the summation runs over the $N_c$ held out records, $p_i$ is the predicted probability that record $i$ is generated, and $y_i$ equals one for generated records and zero otherwise. AUC and balanced accuracy values close to 0.5, together with Brier scores close to 0.25, indicate that complete generated records are difficult to separate from benchmark records.

Table 5. Classifier and personal consistency diagnostics

| ***Diagnostic*** | ***Result*** |
|---:|:---:|
| *Joint patterns used by the classifier* | 448 |
| *Area under the receiver operating characteristic curve* | 0.5272 |
| *Balanced accuracy* | 0.5171 |
| *Brier score* | 0.2492 |
| *Missing common fields* | 0 in both datasets |
| *Invalid common codes* | 0 in both datasets |
| *Child employment industry flag* | 0.660% generated; 0.474% benchmark |

The generated records contain no issues in fields that are available only in the generated dataset. The non-adult employment industry flag occurs in 55,074 generated records and 7,915 benchmark records, corresponding to 0.660% and 0.474% of the two datasets. Since the same pattern occurs in the benchmark, it is treated as a soft plausibility indicator rather than a hard logical violation. The generated rate is 0.186 percentage points higher, which suggests that age and industry combinations remain an area for further calibration.

#### *4.4.4. Household member associations*

Household member associations are often overlooked in recent population synthesis studies. Because our synthetic population is generated from household-personal combined records and associated DAGs, we verify that the output resembles the household structures in the input data.

We evaluate 2-person and 3-person households, sorting members by age to avoid mismatching. We first compare the joint distribution of member age combinations and then expand to age, gender, and race. Figure 14 shows that the R-squared level changes little when more attributes are included, indicating that the general pattern of member associations is robustly captured.

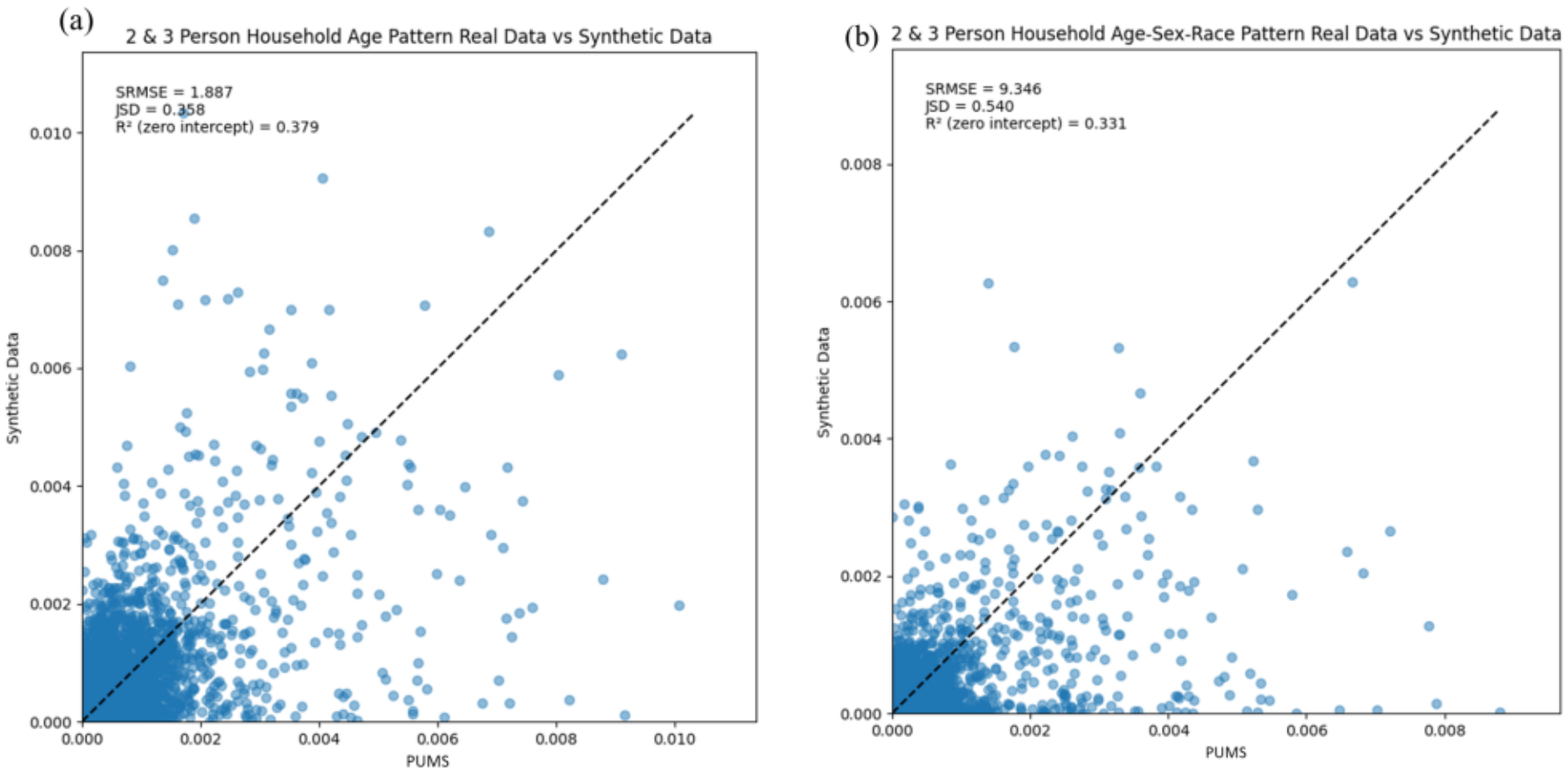

Figure 14. Joint distribution matching for combination of 2 and 3-person household members' (a) age and (b) gender.

To verify that the framework does not yield socio-demographically implausible households, we conduct a series of sanity checks. Across all synthetic households, from single-person to five-person, no "minor-only" households are generated, suggesting that the model internalized the absence of such patterns in the training data as a hard constraint.

Table 6 summarizes the household level validation of the generated two person and three person households against the deterministic 20% PopGen benchmark samples. The marginal distributions are reproduced almost exactly, with JSD values below 0.0011 for geography and household income. Association preservation is measured with Cramér's V: the mean absolute differences range from 0.0148 to 0.0604 for cross member and household to person pairs, indicating that most dependency structures in the benchmark are retained. The classifier two-sample tests yield AUC values of 0.5492 and 0.5364 with balanced accuracy close to 0.5, meaning that complete generated household records are nearly indistinguishable from benchmark records.

At the same time, 89.02% and 99.48% of complete generated patterns are absent from the sampled benchmark, quantifying the diversity added beyond the observed sample, while no invalid geographic codes or unsupported categories occur. The remaining plausibility flag is the share of children with a nonzero employment category, 1.16% and 7.32%, which identifies a specific target for further calibration.

Table 7 reports the cross member Cramér's V comparison for two person and three person households. Most member to member dependencies are reproduced closely: the age associations differ by at most 0.0213, and the employment and English proficiency pairs by less than 0.033. The main exception is the sex association between the first two members, which is underestimated by 0.2060 in two person households and 0.1137 in three person households. This reflects the absence of explicit relationship roles in the data construction, as spousal pairings drive this association in the benchmark.

Table 6. Summary of household level validation against the 20% PopGen benchmark samples

| *Metric* | *Two person households* | *Three person households* |
|---|---|---|
| *Generated households* | 954,537 | 534,023 |
| *20% PopGen benchmark households* | 191,461 | 106,986 |
| *Geography JSD* | 0.0004 | 0.0007 |
| *Household income JSD* | 0.0011 | 0.0010 |
| *Mean cross member absolute difference in V* | 0.0604 | 0.0183 |
| *Mean household to person absolute difference in V* | 0.0148 | 0.0303 |
| *Classifier AUC* | 0.5492 | 0.5364 |
| *Classifier balanced accuracy* | 0.5274 | 0.5023 |
| *Complete patterns absent from sampled benchmark* | 89.02% | 99.48% |
| *Geography values absent from sampled benchmark* | 0 | 0 |
| *Rows with categories outside sampled support* | 0 | 0 |
| *Children with nonzero employment category* | 1.16% | 7.32% |

*Note: JSD denotes Jensen Shannon distance. V denotes Cramér's V. The mean absolute differences are calculated from the association pairs reported in the validation summaries.*

Table 7. Cross member Cramér's V comparison

| *Household size* | *Member pair* | *Generated V* | *PopGen V* | *Absolute difference* |
|---|---|---|---|---|
| *Two-person* | Sex 1 and sex 2 | 0.3125 | 0.5185 | 0.2060 |
| *Two-person* | English proficiency 1 and 2 | 0.1021 | 0.1345 | 0.0324 |
| *Two-person* | Employment 1 and employment 2 | 0.1429 | 0.1457 | 0.0028 |
| *Two-person* | Age 1 and age 2 | 0.4215 | 0.4209 | 0.0005 |
| *Three-person* | Sex 1 and sex 2 | 0.4419 | 0.5557 | 0.1137 |
| *Three-person* | Age 1 and age 3 | 0.2665 | 0.2452 | 0.0213 |
| *Three-person* | Employment 2 and employment 3 | 0.1109 | 0.0949 | 0.0160 |
| *Three-person* | Employment 1 and employment 2 | 0.1324 | 0.1171 | 0.0153 |
| *Three-person* | English proficiency 2 and 3 | 0.1753 | 0.1898 | 0.0146 |
| *Three-person* | Employment 1 and employment 3 | 0.0808 | 0.0707 | 0.0101 |
| *Three-person* | Sex 1 and sex 3 | 0.0542 | 0.0457 | 0.0086 |
| *Three-person* | Sex 2 and sex 3 | 0.0185 | 0.0118 | 0.0067 |
| *Three-person* | English proficiency 1 and 2 | 0.1120 | 0.1055 | 0.0065 |
| *Three-person* | Age 2 and age 3 | 0.2485 | 0.2515 | 0.0031 |
| *Three-person* | English proficiency 1 and 3 | 0.0395 | 0.0424 | 0.0029 |
| *Three-person* | Age 1 and age 2 | 0.3644 | 0.3634 | 0.0010 |

Table 8 evaluates representative higher order joint distributions. Combinations with a moderate number of categories are reproduced well: the four attribute member level combinations and the vehicle, income, and employment combinations have JSD values between 0.0138 and 0.0433, with absent rates below 4%. Accuracy degrades once geography enters the combination, with JSD rising to 0.2843 and 0.3800 and absent rates of 30.51% and 41.60%. This degradation reflects the sparsity of the benchmark in a joint space exceeding 130,000 categories rather than a failure of the lower order distributions, which remain closely matched.

Table 8. Representative higher order joint distribution validation

| *Household size* | *Joint combination* | *JSD* | *SRMSE* | *Absent rate* |
|---|---|---|---|---|
| *Two-person* | Member 1 age, sex, English proficiency, employment | 0.0140 | 0.3744 | 0.39% |
| *Two-person* | Member 2 age, sex, English proficiency, employment | 0.0138 | 0.4032 | 0.53% |
| *Two-person* | Both members age and employment | 0.0898 | 1.1932 | 7.05% |
| *Two-person* | Vehicle, income, member 1 employment | 0.0166 | 0.4456 | 0.55% |
| *Two-person* | Geography, income, member 1 age, employment | 0.2843 | 1.6380 | 30.51% |
| *Three-person* | Member 2 age, sex, English proficiency, employment | 0.0248 | 0.4984 | 1.11% |
| *Three-person* | Member 3 age, sex, English proficiency, employment | 0.0433 | 0.7526 | 3.66% |
| *Three-person* | Employment of all three members | 0.0658 | 0.9576 | 4.49% |
| *Three-person* | Vehicle, income, member 1 employment | 0.0228 | 0.5154 | 0.64% |
| *Three-person* | Geography, income, member 1 age, employment | 0.3800 | 1.6649 | 41.60% |

Taken together, the expanded validation provides stronger evidence that the generated households reproduce the benchmark marginals and most cross member and household to person associations. The remaining differences are concentrated in the sex relationship between the first two members and in sparse combinations that jointly include geography, income, age, and employment. The higher child employment rate in three person households also requires additional examination. These results support the household association objective while defining the aspects that should be improved in later model development.

### *4.4.5. Validation using 2019 census data*

To evaluate robustness across socio-demographic contexts, we perform a cross-temporal validation by synthesizing the 2019 New York City population with the model trained on 2021 data, using 2019 census attributes of residential area, age, and race as conditional inputs.

Updating the conditional inputs to the 2019 ground-truth values of residential location, age, and racial composition tests the model's capacity for "population backcasting". The resulting synthetic population is benchmarked against the 2019 PUMS to quantify predictive accuracy and to confirm that the model captured persistent structural relationships rather than overfitting a single temporal snapshot.

Figure 15 compares the synthetic and benchmark joint probabilities for single-person, two-person, and three-person households over household income, member age, sex, employment status, and residential location.

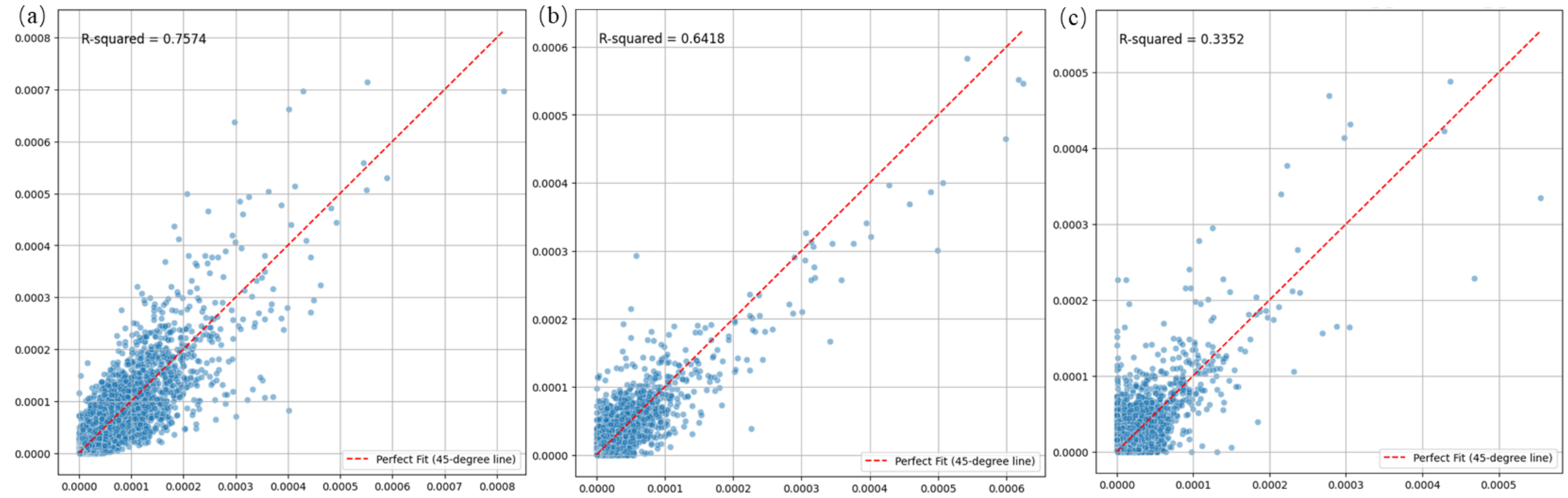

Figure 15. Goodness-of-fit comparison of benchmark vs. synthetic joint distributions for socio-demographic attributes for (a) 1-person, (b) 2-person, and (c) 3-person households.

The model achieves an R-squared of 0.7574 for single-person households (Figure 15(a)), where the attribute space is relatively constrained. Accuracy decreases with household size, to 0.6418 for two-person (Figure 15(b)) and 0.3352 for three-person households (Figure 15(c)). This degradation reflects the curse of dimensionality, as multi-person households add a combinatorial explosion of intra-household relationships. Nevertheless, an R-squared of 0.3352 for three-person households aligns with the performance reported by Borysov et al. (2019) and Garrido et al. (2020), indicating a competitive level of representativeness even within high-occupancy units. Sanity checks on the 2019 output confirm robustness under the shifted conditional inputs: consistent with the 2021 validation, no “minor-only” households are observed.

We also evaluate households containing employed minors, which appear in roughly 0.001% of 2019 PUMS households and about 0.01% of synthetic households, a consistent tenfold expansion. Because only the conditional inputs change between 2021 and 2019, this stable trend indicates that the learned dependencies persist as the marginal distributions shift over time.

### 4.5. Diversity analysis of the synthetic population

Having validated the method, we examine the diversity of the synthesized population: beyond matching marginals and household structures, the synthetic households should also be more diverse than the input sample.

We employ the entropy formulation (Hien & Hoai, 2006) using Eq. (8) to quantify the diversity level of the households in both the PUMS and the synthetic population.

$$E(X) = -\sum x_i \log{(x_i)} \qquad (8)$$

Here, $x_i$ represents the percentage of a specific household group represented by a unique combination of household member attributes. We analyze 2-person and 3-person households, first combining member age, sex, and race. The diversity score is 33.82 for PUMS, and for the synthetic population the score increases to 37.16 (+9.9% vs PUMS). When adding the household attribute of income and vehicle ownership to the combination, the diversity scores are 39.91 and 46.68 (+17.0% vs PUMS) for PUMS and the synthetic population, respectively. By using our proposed methodology, we can increase diversity in the population by 10-17% relative to the PUMS sample data.

We then compare the diversity score using the same attribute selections between the two sets of NYC synthetic population generated from Popgen alone and the proposed framework. Combining member and household level attributes, the score is 20.34 for the Popgen population and 23.02 (+13.2%) for the proposed framework, indicating a more diverse synthesized household structure, which could play an important role in household specific analysis.

## 5. Discussion

### 5.1. Comparison with deep generative household synthesis studies

To position our results within the state of the art, we compare the outcomes obtained in this study with the outcomes reported by four recent deep generative household synthesis studies. Table 9 summarizes the generative core and household representation, the application scale, the benchmark and baseline methods, and the headline validation results reported in each study, together with the corresponding results of this study. Direct numerical comparison should be interpreted with caution because the studies differ in study areas, attribute sets, category resolutions, benchmark definitions, and evaluation protocols. In particular, SRMSE, JSD, and R-squared values are sensitive to the number of joint categories being evaluated, and our validation includes a census tract level geographic attribute with 2,313 categories, which greatly enlarges the joint space relative to the cited studies.

Several observations emerge from Table 9. First, the very low SRMSE values reported by Kukić et al. (2024) correspond to a compact space of three attributes without a geographic dimension, which illustrates how strongly the attribute space affects the reported metrics. Our member level four-attribute SRMSE values of 0.374 to 0.753 are computed over combinations of age, sex, English proficiency, and employment, and they are of the same order as the joint household-individual SRMSE of 0.36 to 0.49 that Sané et al. (2025) report for a three-attribute combination in a much smaller region. The bivariate SRMSE of about 0.84 reported by Aemmer and MacKenzie (2022) for Washington State similarly shows that even two way combinations remain difficult for both generative and traditional synthesizers at realistic category resolutions. Seen in this light, the results of this study are within the range reported by comparable studies while being obtained at a substantially larger scale, with a finer geographic resolution, and with the full set of household member blocks generated jointly.

Table 9. Validation results reported by recent deep generative household synthesis studies and the corresponding results of this study

| *Study* | *Generative core and household representation* | *Application and scale* | *Benchmark and baseline methods* | *Reported headline results* |
|---|---|---|---|---|
| *Aemmer and MacKenzie (2022)* | Linked household VAE and individual CVAE; households generated first, individuals conditioned on household attributes | Washington State, US; 100,000-person test population | Complete state PUMS treated as the target population; PopulationSim (list balancing) as the traditional baseline | Univariate SRMSE 0.293 vs. 0.316 for PopulationSim and bivariate SRMSE 0.843 vs. 0.863 at 100,000 training samples (7.1% and 2.3% reductions) |
| *Kukić et al. (2024)* | One-step Gibbs sampling of complete household records; one sampler per household size; members sorted by decreasing age | Swiss Mobility and Transport Microcensus | Generated data validated against the 2015 microcensus sample with bootstrap confidence intervals; two-step Gibbs sampling and IP as baselines | Average SRMSE of 0.01 (first order), 0.03 (second order), and 0.05 (third order) for age, gender, and household size; attribute-level SRMSE of 0.008-0.131 |
| *Sané et al. (2025)* | SVAE-Pop2 (single VAE with padded records) and MVAE-Pop2 (one VAE per household size) | Lyon area, France; about 1.4 million persons and 650,000 households | Reference population of the study area used as test data; the two designs benchmarked against each other | Individual marginal SRMSE 0.15-0.17; household marginal SRMSE 0.21-0.22; joint household-individual combination (household size, age, sex) SRMSE 0.36-0.49 with correlation 0.93-0.96 |
| *Qian et al. (2026)* | TL-VAE on restructured wide-format household records with member blocks; transfer learning to census tract marginals | Delaware (1million population) and North Carolina (11.3 million population), US | Self-similarity protocol: 5% microdata subsample for training, remaining PUMS as evaluation benchmark; IPF and copula variants (BN, TVAE, CTGAN) as baselines | Contingency similarity 0.946; α-precision 0.981; β-recall 0.732; sampling zeros coverage rate 0.172; KL divergence 0.016 |
| *This study* | ciDATGAN with selected DAGs and an IPF-anchored conditional population; joint household record with repeated member blocks; one model per household size up to five members | New York State, US; nearly 20 million persons and 7.5 million households | Models trained on 40% of records; evaluated against 100% PUMS joint distributions and deterministic 20% PopGen holdout samples; PopGen as the traditional baseline | Joint distribution of PUMA, age, race, and disability with R-squared of 0.602 and JSD of 0.280; member-level four-attribute combinations with JSD of 0.014-0.043 and SRMSE of 0.374-0.753; classifier two-sample test AUC of 0.527-0.549; diversity gain of 10-17% over PUMS and 13.2% over PopGen |

*Note: Results are quoted as reported in the cited studies. Values are not directly comparable across studies because the study areas, attribute sets, category resolutions, benchmarks, and evaluation protocols differ.*

Second, recent studies increasingly describe the quality of synthetic populations in terms of fidelity, diversity, and generalizability (Qian et al., 2026). Under this terminology, our classifier two-sample tests, with AUC values between 0.527 and 0.549, indicate high fidelity, since complete generated records are nearly indistinguishable from benchmark records. Our entropy based diversity gains of 10-17% over PUMS and 13.2% over PopGen, together with the high share of complete household patterns absent from the sampled benchmark, correspond to the generation of sampling zeros, that is, plausible attribute combinations that are missing from the sample. Qian et al. (2026) report a sampling zeros coverage rate of 0.172 and a β-recall of 0.732 for their TL-VAE, and Garrido et al. (2020) document the same tradeoff between recovering sampling zeros and avoiding structural zeros. Our plausibility checks, which found no minor-only households but an elevated child employment indicator, show the same tension: added diversity must be balanced against the risk of generating implausible records.

The relation between each generative model and its traditional baseline is also consistent across studies. Aemmer and MacKenzie (2022) report SRMSE reductions of 7.1% and 2.3% relative to PopulationSim, and Qian et al. (2026) report a substantially higher β-recall than IPF (0.732 against 0.431) at nearly identical α-precision, indicating that the main advantage of deep generative models over traditional synthesizers lies in coverage and diversity rather than marginal accuracy. Our results reproduce this pattern: the framework raises household entropy by 13.2% relative to PopGen and generates a large share of household patterns absent from the sample, while keeping marginal JSD values below 0.0013 and classifier separability near chance.

Last but not least, the comparison also highlights the different mechanisms that the literature uses to keep flexible generators aligned with known population structure. Qian et al. (2026) decouple dependency learning from marginal alignment through transfer learning. Kukić et al. (2024) embed consistency in the design of the conditional distributions of the Gibbs sampler. Sané et al. (2025) rely on household size specific submodels, which our framework shares. Our framework instead supplies the alignment through an externally prepared, IPF anchored conditional population, so that the generative model is constrained by a fixed skeleton of residence area, age, and race. This anchored hybrid design attains a fidelity and diversity balance comparable to fully deep learning based approaches while retaining the explicit control over geographic totals that traditional methods provide.

### 5.2. Limitations and future work

Several limitations of this study should be acknowledged, and each points to a direction for future work. First, a direct experiment using the same records and data split with another household level generative model, such as a conditional variational autoencoder, was not conducted in this study. The comparison in Section 5.1 is therefore literature based, and no claim of empirical superiority over existing household generative models is made. A controlled benchmark of ciDATGAN based, VAE based, and Gibbs sampling based household synthesis on a common dataset with shared evaluation metrics is a priority for future research.

Second, household type and relationship roles are not modeled. Household members are ordered by age, and no explicit spouse or parent-child constraints are imposed, so role dependent structures are only captured implicitly. This is the most likely cause of the underestimated sex association between the first two household members, where the Cramér's V difference reaches 0.206 for two person households. Future work will incorporate household type or relationship roles as attributes, following Kukić et al. (2024), or introduce role aware constraints in the data construction stage.

Third, household consistency is learned implicitly rather than enforced by explicit rules. The validation shows that hard structural patterns, such as the absence of minor-only households, are internalized by the model, but soft implausibilities remain: individuals under 18 years old with a nonzero employment category account for 1.16% of two person households and 7.32% of three person households, and sparse combinations of geography, income, age, and employment reach JSD values of up to 0.386. Future work will add explicit constraint screening, rejection rules, or post hoc calibration to control such combinations without sacrificing diversity.

Fourth, the current application contains at most 48 attributes in one household record. It does not establish the performance of the framework for datasets containing hundreds of variables. Such an application would require additional testing of computational cost, network complexity, sample sparsity, and synthetic data quality. Similarly, households with six or more members are completed by replicating sample records rather than generated, and although they account for less than 5% of households, padded single model designs such as the SVAE-Pop2 of Sané et al. (2025) offer a possible route to cover large households generatively.

Fifth, the DAG selection relies on cross validated AIC as the single screening criterion, and the focused edges reflect expert judgment. Alternative criteria such as BIC, held out likelihood, or Bayesian scores, sensitivity analyses of the expert edge set, and more advanced structure learning methods that reduce DAG connectivity are all worth exploring in future research. Moreover, low dimensional visual diagnostics such as principal component analysis or t-distributed stochastic neighbor embedding were not used as primary validation measures because most attributes are categorical and such projections are sensitive to encoding and tuning choices; they could be added as supplementary visualizations with observed and synthetic records projected jointly under a common encoding.

Finally, the training data span 2017 to 2021 and therefore mix the periods before and after the COVID-19 pandemic, and the temporal transfer of the trained models was tested only for the 2019 backcast. Future work will examine transferability across additional regions and years and integrate the released synthetic population with downstream agent based simulation and equity analyses.

## 6. Conclusion

Synthetic populations are increasingly important in urban and transportation studies, where higher resolution household and individual information supports behavioral and equity evaluations, such as the ongoing equity analysis by the SEMPACT center (Bian et al., 2024) is a typical use case. However, generating synthetic population can pose several challenges. Classic methods such as IPF generate reliable results, but they cannot produce out-of-sample observations. Later probabilistic approaches can in principle produce unobserved combinations. However, the diversity of the generated population remains limited because the fitted distributions are tied to the observed sample, and neither family of methods remains tractable when the joint attribute space becomes very large.

Deep learning provides a new way for population synthesis but offers less control over the generation process. In this study, we propose a framework that generates household and individual level attributes in a more controlled manner. The framework builds on the existing ciDATGAN model as its generative core and combines it with PopGen, Bayesian network structure learning, ordinary least squares, and random forest procedures. The deterministic PopGen synthesis provides the anchor that fixes dependable attributes and geographic totals, while the DAG regularized ciDATGAN supplies the flexibility to learn the remaining household and member relationships. Household member attributes are combined with household level attributes in the input sample so that member associations can be extracted directly, and multiple methods are used to generate the DAGs. The DAG with the best mean cross validated AIC score is selected as the conditional dependency structure used to regularize ciDATGAN. The selected graph is not interpreted as a causal model. Popgen further generates the partial population used as the conditional input, adding another layer of control.

We apply the framework to generate a synthetic population for the whole of New York State, containing nearly 20 million individuals and 7.5 million households. Household-personal combined tables are prepared per household size, capturing member associations for households of up to 5 people. The synthetic population matches census marginals well, generally matches individual and household level joint distributions, and reproduces the member associations present in the input sample.

This study provides a systematic approach to generating synthetic populations with both individual and household level information, handling large multivariate datasets in a controlled manner while capturing member associations. Despite multiple training processes, the overall runtime is shorter than that of PopGen, and the efficiency gain grows with dimensionality.

The limitations of this study and the corresponding directions for future research are discussed in Section 5.2. Overall, the results indicate that anchoring a flexible deep generative model with a deterministic synthesis skeleton is a practical route to producing large scale synthetic populations that preserve household structure while adding diversity beyond the observed sample.

## Acknowledgments

This work was supported by the SEMPACT University Transportation Center. Hongying Wu was funded by the GSET program at NYU Tandon School of Engineering, and Linfei Yuan was funded by the Undergraduate Summer Research Program at NYU.